\documentstyle[aps,prb,epsfig]{revtex}

\begin{document}



\title{Deformation potential scattering from dislocations in III-V nitride quantum wells}

\author{Debdeep Jena\thanks{email: djena@engineering.ucsb.edu} and U. K. Mishra}
\address{Department of Electrical and Computer Engineering\\
                University of California, Santa Barbara \\
            CA, 93106}
\maketitle

\begin{abstract}
We present a theory of deformation potential carrier scattering of
two-dimensional electron gases from the strain fields surrounding
dislocations.  The results are applied to study the transport
characteristics in III-V nitride two-dimensional electron gases in
Al(Ga)N/GaN quantum wells.  A hypothetical charged core has been
associated as the {\em only} scattering potential for analyzing
experimental results for transport studies of the nitrides; we
critically examine this assumption in light of our results of
strain field scattering.  By computing the effect of all possible
scattering mechanisms we gauge the importance of strain field
scattering from dislocations.
\end{abstract}



%

Many years ago, the effect of `cold working' on metallic
resistivity was studied in detail
\cite{koehler},\cite{mack_son},\cite{landauer},\cite{dexter}. Cold
working is a technique of introducing controlled amount of
dislocations by deformation; the study showed that metallic
conductivity is reduced by scattering of conduction electrons from
strain fields that develop around dislocations.  The effect of
strain fields on electronic energy levels and charge transport in
semiconductors is a widely studied topic, assuming special
importance in the problems of lattice scattering and optical
transitions in strained heterostructures.

Localized strain fields exist around point and extended defects in
semiconductors.  Traditionally in electronic transport theory one
considers charge scattering by coulombic interaction of mobile
carriers with charged defects; strain fields associated with
defects is generally neglected.  This approximation is justified
for substitutional donors/acceptors for example, since the lattice
distortion around them is minimal.  However, as our work shows for
dislocations, which may or may not be charged, the strain fields
can contribute substantially to scattering of mobile carriers in
semiconductors, just as in metals.  Electron-strain field
interaction will affect transport properties for
vacancies/interstitials as well; we do not consider them in this
work.

Dislocation scattering effect on two-dimensional electron gas
(2DEG) transport in AlGaN/GaN heterostructures has been recently
studied assuming coulombic scattering from a charged dislocation
core \cite{myfirstpaper}.  In this work, we solve the general
problem of the effect of scattering from the strain field
surrounding edge dislocations for a 2DEG. We examine it's
importance by applying the results for the AlGaN/GaN system.  It
is important to note that this form of scattering arises {\em even
if the dislocation core is uncharged}.

Dislocations set up a strain field around them with atoms
displaced from their equilibrium positions in a perfect crystal.
The band extrema (conduction band CB minimum, valence band VB
maximum) shift under influence of the strain fields.  The
magnitude of spatial variation of the band extrema to linear order
in strain is given by the deformation potential theorem of Bardeen
and Shockley \cite{bardeen_shockley}.

We start with a suitable model for behavior of quantum well
band-edges in the presence of a localized strain field, such as
around a dislocation.  We assume a flat quantum well, with no
built in fields, which houses a 2DEG\cite{comment_built_in_field}.
Our work deals with electron transport; the problem of hole
transport can be formulated in a similar fashion.  The effect of a
strain in the quantum well is to shift the conduction and valence
band edges.  The shift in the conduction band edge was shown by
Chuang\cite{chuang} to be

\begin{equation}
\Delta E_{C} = a_{C} \hspace{0.1cm} Tr(\epsilon)
\end{equation}

where $a_{C}$ is the conduction band deformation potential, and
$Tr(\epsilon)=\epsilon_{xx}+\epsilon_{yy}+\epsilon_{zz}=\delta
\Omega/\Omega$ is the trace of the strain matrix.  The trace is
also equal to the fractional change in the volume of unit cells
($\delta \Omega/\Omega$).

In our model, we assume dislocations with their axes perpendicular
to the quantum well plane.  We also assume that the 2DEG is
perfect, which means there is no $z$ direction spread along the
growth axis.  Considering a realistic 2DEG would require
incorporation of form factors; perfect 2DEG is chosen for
simplicity.  As an electron in the 2DEG approaches a dislocation,
it experiences a potential due the strain around the dislocation,
which causes scattering (see Figure[1] for a schematic).  The
strain distribution radially outward from an edge dislocation is
well known\cite{yu}.  Combined with Equation [1] we get the
necessary perturbing potential responsible for electron scattering

\begin{equation}
\delta V=\Delta E_{C}= a_{C} Tr(\epsilon)= -\frac{a_{C}
b_{e}}{2\pi} \frac{1-2\gamma}{1-\gamma} \frac{sin(\theta)}{r}.
\end{equation}

Here $b_{e}$ is the magnitude of the burger's vector of the edge
dislocation, and $\gamma$ is the Poisson's ratio for the crystal.
$\epsilon_{zz}=0$ for an edge dislocation, and nonzero for a screw
dislocation.  For a screw dislocation in a cubic crystal, the
strain field has purely shear strain, causing no
dilatation/compression of the unit cells.  This means there can be
no deformation potential scattering for screw dislocations in
cubic crystals.  However, for uniaxial crystals such as GaN, the
argument does not hold, and there is a deformation potential
coupling even for screw dislocations for bulk transport.  We limit
ourselves to the simpler case of edge dislocations.

The matrix element of the perturbation for scattering of a 2DEG
electron from state $|{\bf k_{i}}>$ to state $|{\bf k_{f}}>$ is
needed for evaluating scattering rates in the Born approximation.
Position space representations of the states are given by plane
waves

\begin{equation}
<r|k_{i,f}>=\frac{1}{\sqrt{S}}e^{i{\bf k_{i,f} \cdot r}}.
\end{equation}

Here ${\bf k_{i,f}}$ are the 2D wavevectors of the initial (i) and
final (f) states, ${\bf r}$ is the 2D space coordinate, and $S$ is
the macroscopic 2D area.  The wavevectors of the initial and final
states are both perpendicular to the dislocation axis.  The matrix
element $<{\bf k_{f}}|\delta V(r,\theta)|{\bf k_{i}}>$ is given by
the 2D Fourier transform of the scattering potential (the Born
approximation)\cite{ziman}

\begin{equation}
\delta V(q,\phi)=\int e^{i{\bf (k_{i}-k_{f}) \cdot r}} \delta
V({\bf r}) d^{2}r = \frac{b_{e} a_{C}}{2\pi S}
\frac{1-2\gamma}{1-\gamma} \frac{sin(\phi)}{q},
\end{equation}

where $q=|{\bf k_{f}-k_{i}}|$ and $\phi$ is the angle between
${\bf q}$ and ${\bf b_{e}}$, the Burger's vector.  For taking into
account screening of this perturbation by mobile charges, the
matrix element is scaled by the Lindhard dielectric function in
the long-wavelength limit $\epsilon(q)=1+\frac{q_{TF}}{q}$, where
$q_{TF}=\frac{2}{a_{B}^{*}}$ is the Thomas-Fermi wavevector
($a_{B}^{*}$ is the effective Bohr radius in the semiconductor).
Summing the square of the matrix element over all scatterers in
the dilute scatterers limit requires an average of the angular
dependence over random orientations of the burger's vectors for
different dislocations; averaging yields
$<sin^{2}(\phi)>=\frac{1}{2}$.  Transport scattering rate is found
by Fermi's golden rule; for scattering into the single final state
${\bf k_{f}}$, the rate is given by

\begin{equation}
\frac{1}{\tau}=\frac{2\pi}{\hbar}|\delta V(q)|^{2} \delta(E_{\bf
k_{i}}-E_{\bf k_{f}}),
\end{equation}

where $\tau$ is the scattering rate, $\hbar$ is the reduced
Planck's constant, and the $\delta$ function is a statement of the
elastic nature of scattering, conserving energy between the
initial ($E_{{\bf k_{i}}}$) and final ($E_{{\bf k_{f}}}$) states.

To find the ensemble rate, we sum over all the available final
states in the 2D density of states, and evaluate the transport
scattering rate \cite{davies}

\begin{equation}
\frac{1}{\tau_{tr}}=\frac{N_{disl}m^{*}b_{e}^{2}a_{C}^{2}}{2\pi
k_{F}^{2}\hbar^{3}} (\frac{1-2\gamma}{1-\gamma})^{2}
\underbrace{\int_{0}^{1}
\frac{u^{2}}{(u+\frac{q_{TF}}{2k_{F}})^{2}\sqrt{1-u^{2}}}du}_{I(n_{s})}.
\end{equation}

Here, $N_{disl}$ is the 2D density of threading edge dislocations,
$m^{*}$ is the effective mass of conduction electrons in the 2DEG,
and $k_{F}=\sqrt{2 \pi n_{s}}$ is the Fermi wavevector ($n_{s}$
being the 2DEG electron sheet density).

The dimensionless integral $I(n_{s})$ is dependent only on the
sheet density $n_{s}$, and can be evaluated explicitly.  Since the
expression is long and does not contain any extra information, we
plot the dependence of the integral factor on the sheet density in
Figure[2].  Finally, we arrive at the dislocation strain field
scattering limited electron mobility given by the Drude result
$\mu=e\tau_{disl}^{strain}/m^{*}$

\begin{equation}
\mu_{disl}^{strain}=\frac{2e\hbar^{3} \pi
k_{F}^{2}}{N_{disl}{m^{*}}^{2}b_{e}^{2} a_{C}^{2}}
(\frac{1-\gamma}{1-2\gamma})^{2} \frac{1}{I(n_{s})}.
\end{equation}

Quantities needed for a numerical evaluation are the magnitude of
the Burger's vector $b_{e}=a_{0}=3.189$\AA, the conduction
electron effective mass $m^{*}=0.2m_{0}$ ($m_{0}$ is free electron
mass), Poisson's ratio for the crystal, $\gamma=0.3$ \cite{yu},
and the conduction band deformation potential $a_{C}$.

For uniaxial crystals such as the wurtzite crystal, the second
rank deformation potential tensor $\Xi_{ij}$ has two independent
components, $\Xi_{1}$ and $\Xi_{2}$ at the $\Gamma$ point in the
E-k diagram.  The volume change (compression or dilatation) leads
to a shift in the band gap

\begin{equation}
\Delta E_{G}= \Xi_{1}\epsilon_{zz} +
\Xi_{2}(\underbrace{\epsilon_{xx}+\epsilon_{yy}}_{\epsilon_{\perp}})
\end{equation}

Where $\Xi_{1}=a_{1}=-6.5$eV and $\Xi_{2}=a_{2}=-11.8$eV for
GaN\cite{vurga}.  For an edge dislocation, there is no strain
along the $z$ (0001) axis ($\epsilon_{zz}=0$); thus only $\Xi_{2}$
will be required in our analysis.  The deformation potential has
contributions from both the CB and the VB,
$\Xi_{2}=\Xi_{2}^{CB}+\Xi_{2}^{VB}$.  We require only the
conduction band deformation potential for our calculation.
Separate experimental values of the conduction and valence band
deformation potentials are not available for GaN at present. We
use an approximation of $a_{C}=\Xi_{2}^{CB}=-8.0$eV (and
$\Xi_{2}^{VB}=-3.8$eV) for numerical estimates.  This split in CB
and VB deformation potentials is assumed following the general
trend of other III-V semiconductor deformation potentials.

We now apply the derived results to 2DEGs at AlGaN/GaN
heterojunctions.  To do a comparison of the different scattering
mechanisms, we plot mobility limited by each scattering mechanism
for a range of 2DEG densities in Figure[3].  The total low
temperature mobility is evaluated by Matheissen's rule
($\mu_{tot}^{-1}=\sum_{i} \mu_{i}^{-1}$, $\mu_{i}$ being mobilites
limited by individual scattering mechanisms).  In the same plot,
we plot the highest reported 2DEG mobilities in Al(Ga)N/GaN
2DEGs\cite{manfra},\cite{yulia}. The calculation was done for
background impurity concentration $10^{16}/cm^{3}$, remote
(surface) donor density equal to the 2DEG density\cite{ibbo},
interface roughness characterized by island of height
$\Delta=2.5$\AA and correlation length $L=10$\AA, and charged core
filling factor $f=0.3$\cite{myfirstpaper}.  The barrier alloy
concentration was $9\%$.

It is evident that scattering at high carrier densities is
dominated by alloy scattering, and in the absence of an alloy
barrier, interface roughness scattering.  There is a noticeable
jump in mobility at higher 2DEG densities in passing from an AlGaN
(empty circles) to AlN barrier (filled circles), marking the
removal of alloy scattering.

However, at low carrier densities, dislocation scattering
dominates electron transport properties. Even for an uncharged
dislocation core, the strain field-deformation potential
scattering is large enough to limit low temperature electron
mobilities.  In Figure[4], we plot total mobilities calculated for
three different dislocation densities $N_{disl}=5 \times
(10^{8},10^{9},10^{10})/cm^{2}$.  It is important to note that
dislocation scattering (be it from a charged core or from the
strain field) is much stronger than charged impurity scattering
for impurity and dislocation densities typical in the III-V
nitrides.

In addition to the deformation potential scattering from the
strain fields, there is also a possibility of piezoelectric fields
associated with dislocations in non-centrosymmetric crystals as
GaN.  However, we expect this form of scattering to be
negligible\cite{yu}.  The effect of screw dislocations on
transport in uniaxial crystals is a more subtle question, and we
postpone it for a future work.

In conclusion, we demonstrated that strain fields surrounding
dislocations affect measured electron transport properties in a
2DEG.  We derived scattering rates for deformation potential
scattering from strain fields of edge dislocations.  The
theoretical results were applied to the case of III-V nitride
2DEGs.  By comparison with all low-temperature scattering
mechanisms, the importance of dislocation scattering (originating
from strain fields as well as charged cores) was highlighted.

The authors would like to thank P. Waltereit,  J. Singh, A.
C.Gossard, and J. Speck for useful discussions.  The authors
acknowledge funding assistance from J. Zolper (ONR-IMPACT), C.
Wood (ONR-POLARIS) and G. Witt (AFOSR).

%




\begin{figure}
\caption{The band electron experiences the depicted CB minimum
fluctuation caused by strain fields around an edge dislocation.
Strain is anisotropic, with maximum strain in directions
perpendicular to the Burger's vector. The energy is in arbitrary
units.} \label{fig1}
\end{figure}

\begin{figure}
\caption{Plot of the dependence of the dimensionless integral
$I(n_{s})$ on the 2DEG sheet density $n_{s}$.} \label{fig2}
\end{figure}

\begin{figure}
\caption{Contribution of different scattering processes to low
temperature electron mobility.  There is a good match of
theoretically predicted mobility given by the broken line and the
experimentally measured mobility.  Experimental results for
AlGaN/GaN 2DEGs are plotted as empty circles, and for AlN/GaN
layers as filled circles.  Dislocation scattering is dominant at
low 2DEG densities.} \label{fig3}
\end{figure}

\begin{figure}
\caption{The effect of strain field scattering limited electron
mobility in an Al(Ga)N/GaN 2DEG as a function of the sheet density
for three different dislocation densities, $N_{disl}=5 \times
(10^{8},10^{9},10^{10})/cm^{2}$.} \label{fig4}
\end{figure}

\pagebreak

\thispagestyle{empty}

\begin{figure}[h]
\begin{center}
\leavevmode \epsfxsize=4in
\epsfbox{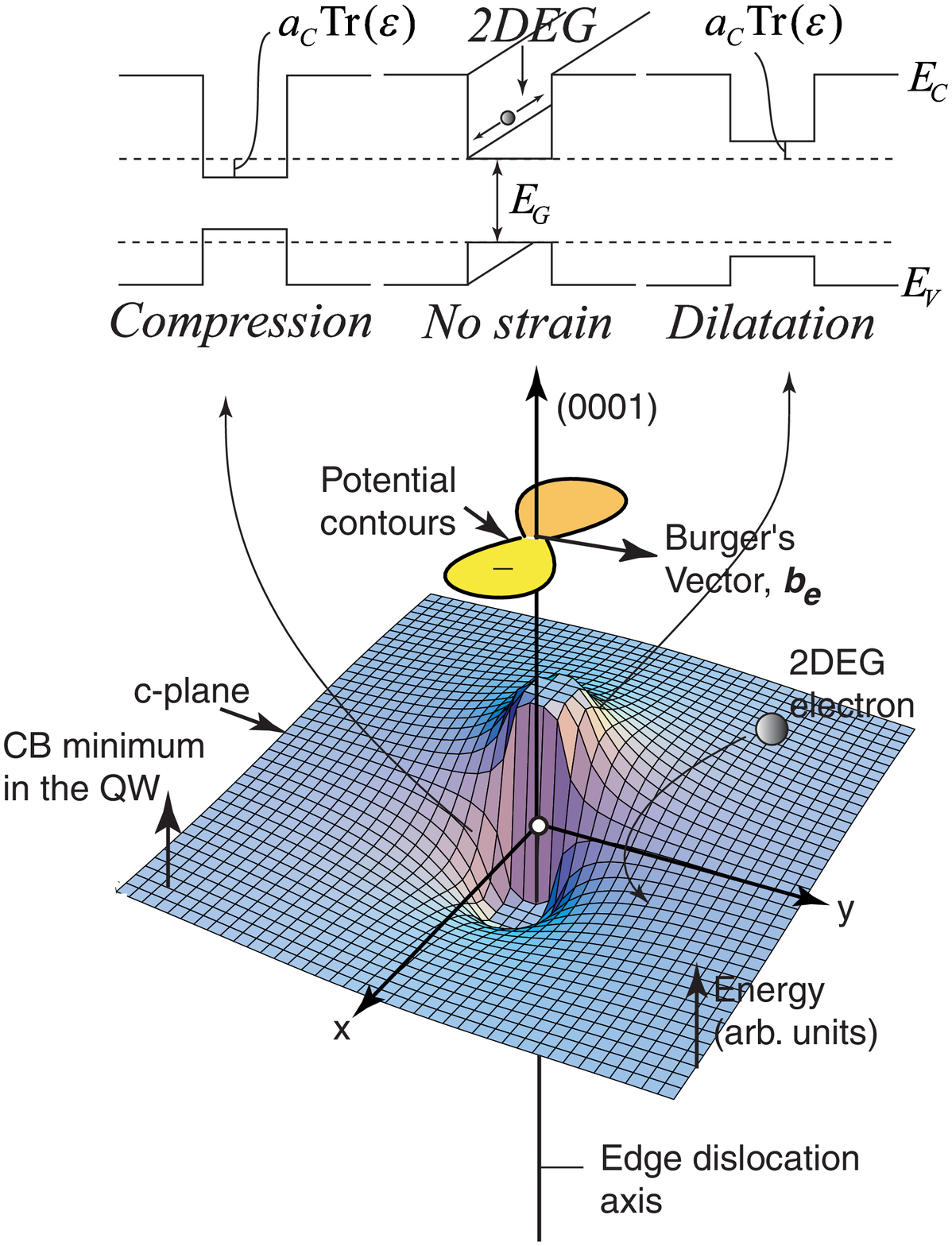}
\end{center}
\end{figure}

\vspace{3in} \noindent {\bf Figure 1\\} {D. Jena and U. K. Mishra}

\pagebreak

\thispagestyle{empty}

\begin{figure}[h]
\begin{center}
\leavevmode \epsfxsize=5in
\epsfbox{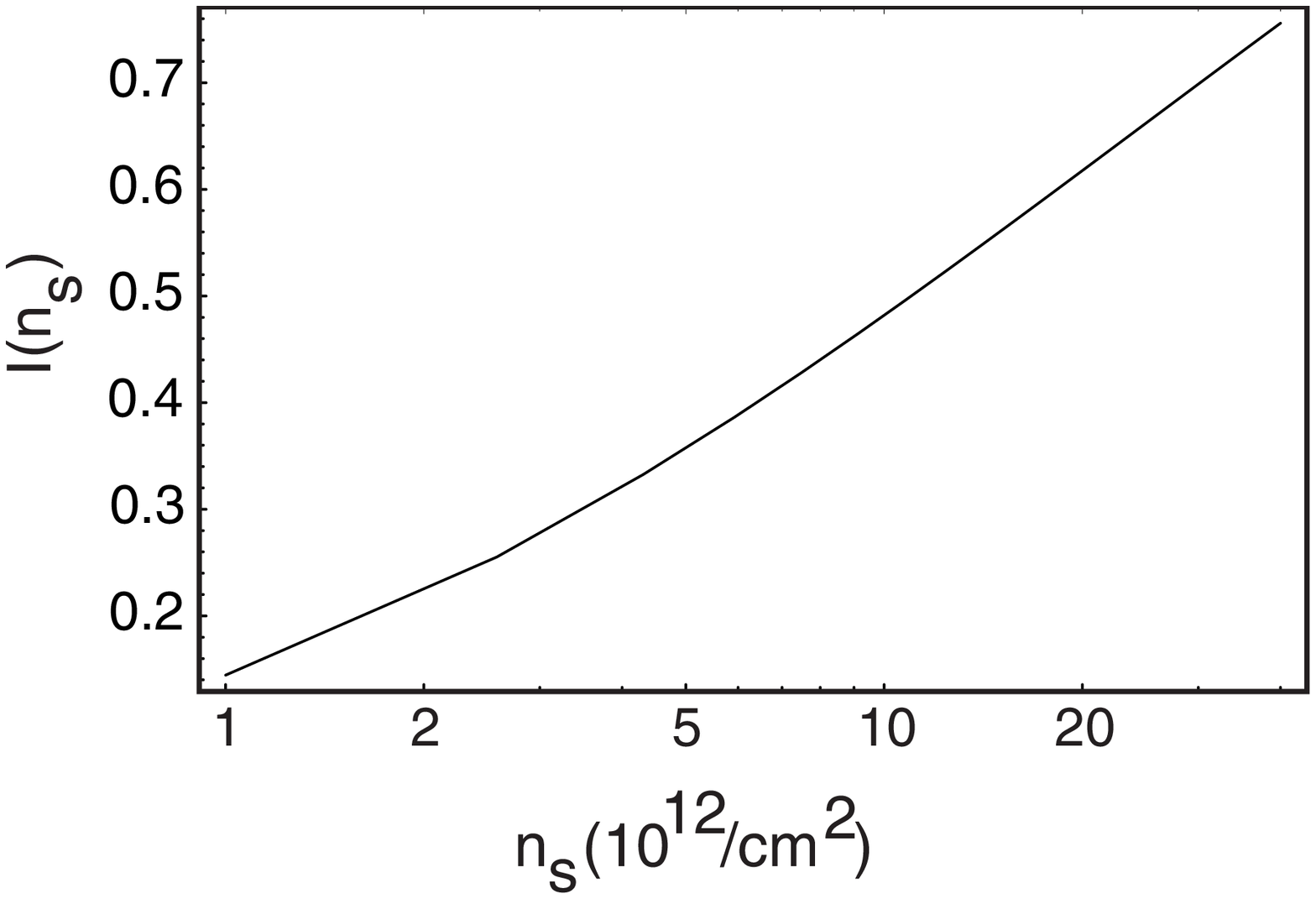}
\end{center}
\end{figure}

\vspace{3in} \noindent {\bf Figure 2\\} {D. Jena and U. K. Mishra}


\pagebreak \thispagestyle{empty}

\begin{figure}[h]
\begin{center}
\leavevmode \epsfxsize=5in
\epsfbox{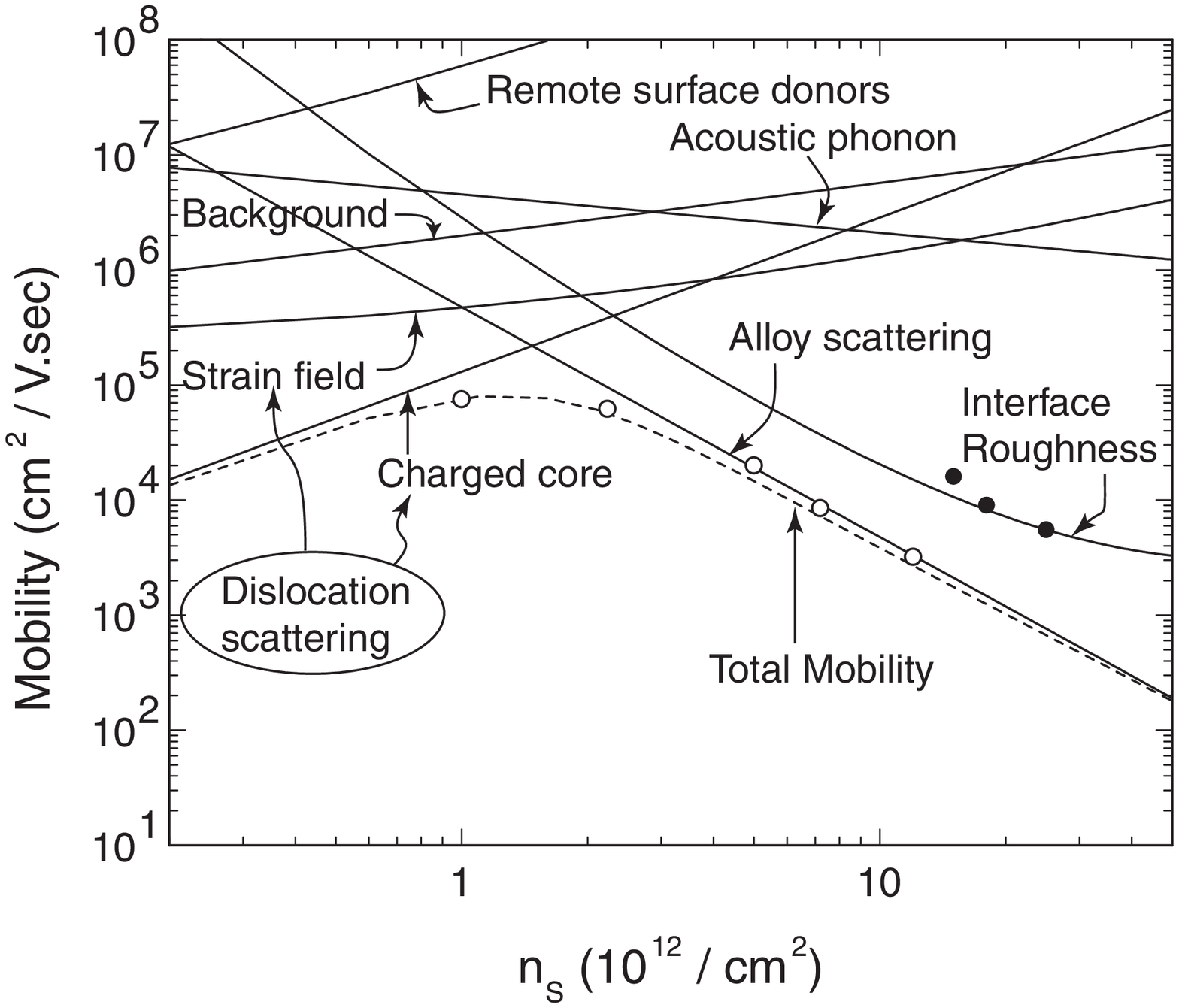}
\end{center}
\end{figure}

\vspace{3in} \noindent {\bf Figure 3\\} {D. Jena and U. K. Mishra}



\pagebreak \thispagestyle{empty}

\begin{figure}[h]
\begin{center}
\leavevmode \epsfxsize=5in
\epsfbox{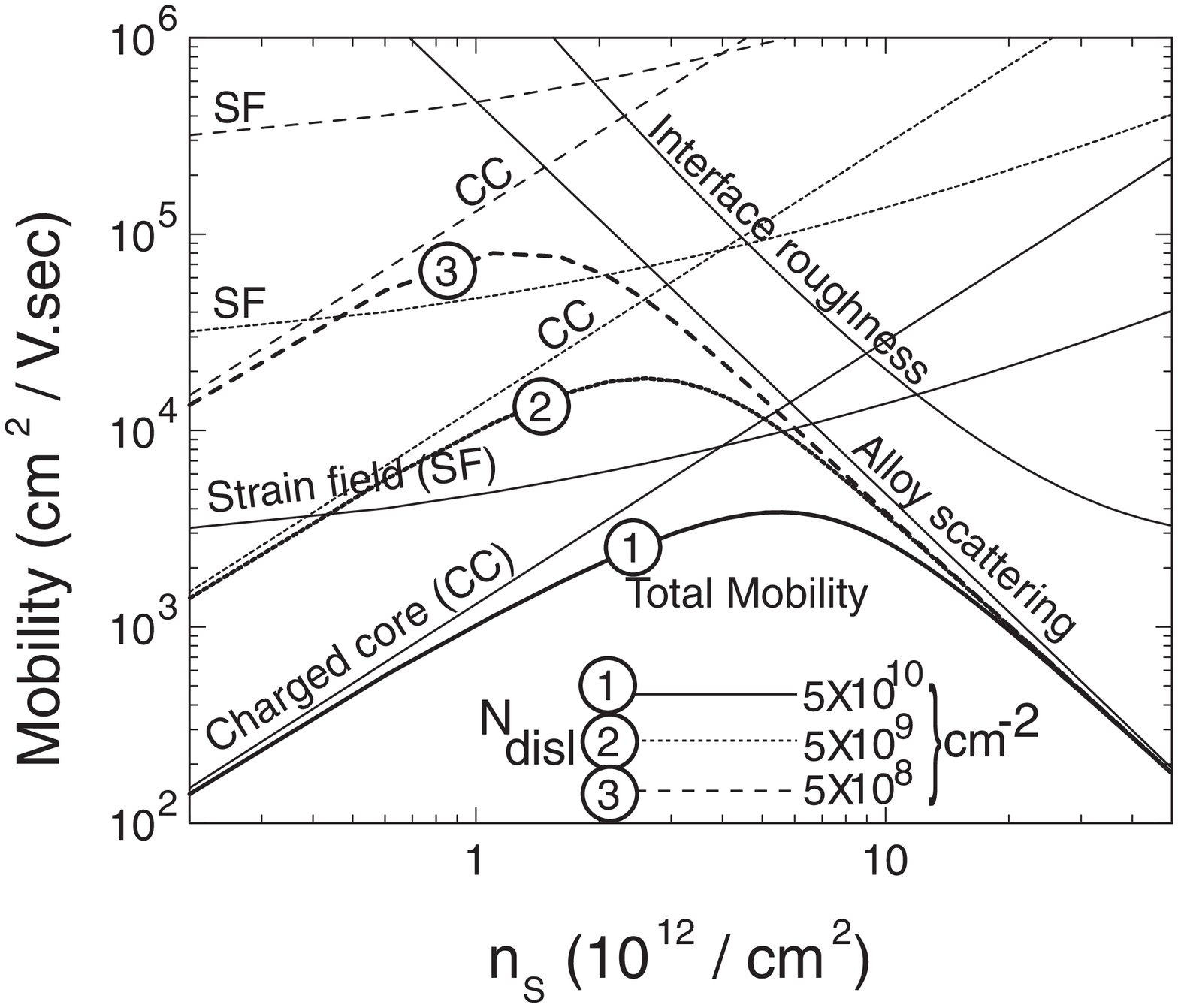}
\end{center}
\end{figure}

\vspace{3in} \noindent {\bf Figure 4\\} {D. Jena and U. K. Mishra}



\begin{references}
\bibitem{comment_built_in_field} This assumption does not hold for
III-V nitrides, which have strong polarization fields in QWs.
However, the assumption introduces negligible error, since we
assume a perfect 2DEG.
\end{references}
\end{document}